\documentclass{webofc}

\usepackage[varg]{txfonts}   
\usepackage{hyperref}
\usepackage{url}
\hypersetup{colorlinks=true,citecolor=blue,urlcolor=blue,linkcolor=blue}
\begin{document}
\title{State-of-the-art of Strangeness in Quark Matter Theory 2024}
%
%

\author{\firstname{Jacquelyn} \lastname{Noronha-Hostler}\inst{1}\fnsep\thanks{\email{jnorhos@illinois.edu}}
}

\institute{Illinois Center for Advanced Studies of the Universe, Department of Physics, 
University of Illinois at Urbana-Champaign, 1110 W. Green St., Urbana IL 61801-3080, USA
}

\abstract{
I discuss the theoretical developments in  related to Strangeness in Quark Matter (SQM) leading up to the SQM2024 conference. These advances include mapping out the Quantum Chromodynamics phase diagram; puzzles that exist in hadron physics from light to heavy particles; and advanced in relativistic hydrodynamics with the inclusion of spin and magnetic fields.  
}
\maketitle
\section{Introduction}
\label{intro}
 I will summarize recent advances in high-energy nuclear theory, pressing questions within the field,  and make connections to astrophysics, gravity, nuclear structure,  and cold atoms.

\section{QCD Phase Diagram}
\label{sec:QCDphases}

The QCD phase diagram cannot be systematically studies in a single experiment or astronomy observation, rather it requires input from heavy-ion collisions HIC (high temperatures $T$, low to intermediate $n_B$, range of charge fractions $Y_Q=Z/A=n_Q/n_B=[0.38,0.5]$), cold neutron stars NS ($T=0$, large $n_B$, $Y_Q\lesssim 0.1$), binary NS mergers ($T>0$, large $n_B$, $Y_Q\sim  [0.01,0.2]$), and low energy nuclear physics ($T\sim 0$, nuclear saturation density $n_B=n_{sat}$, $Y_Q\sim 0.5$). 
These leave gaps in the QCD phase diagram that may be difficult to interpret.  
Theoretical input is important e.g. causality and stability constraints $0\leq c_s^2\leq 1$,  first principle lattice or perturbative QCD (pQCD) calculations, or well-defined effective field theories (EFT) such as chiral EFT. 
Constraints are summarized in \cite{MUSES:2023hyz}.
\emph{Key questions: } How to interpret  new Beam Energy Scan II data? How does QCD matter move at $n_B\neq 0$?   How does HIC/QCD constrain the NS equation of state (EOS)?  Can out-of-equilibrium effects shed light on the core of NS?

{\bf HIC/critical point} The (potential) QCD critical point \cite{Stephanov:1998dy}  separates the boundary between quarks/gluons and hadrons. 
While current first principle calculations cannot determine its existence/location \cite{Bollweg:2022rps}, recent calculations in other theoretical frameworks \cite{Fu:2019hdw,Gunkel:2021oya,Basar:2021hdf,Dimopoulos:2021vrk,Hippert:2023bel,Sorensen:2024mry}  indicate a critical point at $\mu_B/T\sim 4-8$. 
Preliminary measurements exist for net-proton fluctuations $\kappa_4/\kappa_1$ across $\sqrt{s}$ \cite{ZHANGSQM} appearing flat.
Factorial cumulants show qualitative features  predicted from critical point models \cite{Bzdak:2019pkr}.
Viscous effects may smear out or alter critical fluctuations \cite{Dore:2022qyz} and the critical region can either cause critical lensing or minimize the effect \cite{Nonaka:2004pg,Dore:2022qyz,Pradeep:2024cca,Karthein:2024zvs}.
Theorists are developing frameworks (initial state, fluid dynamics, hadron interactions) that dynamically study $n_B\neq 0$ (and strangeness, $n_S\neq 0$, and electric charge, $n_Q\neq 0$). 
The Taylor series EOS from lattice is only valid up to  $\mu_B/T\sim [2-3]$ \cite{Bollweg:2022rps}, which does not cover the entire EOS needed for Beam Energy Scan II. 
An alternative expansion was developed that  improves the reach in $\mu_B/T$ \cite{Borsanyi:2021sxv} and is coupled to 3D Ising \cite{Kahangirwe:2024cny}.
In hydro simulations, every fluid cell must have a valid EOS.  In \cite{Plumberg:2024leb} a solution to a limited EOS was developed in 4-dimensions ($T,\mu_B,\mu_S,\mu_Q$) to smoothly map to a conformal EOS for out-of-bound fluid cells. 
The MUSES collaboration   provides open-source EOS across the phase diagram \cite{MUSESalpha}.

Initial conditions either include baryon stopping through strings \cite{Shen:2017bsr}, a 3-fluid model \cite{Cimerman:2023hjw}, color glass condensate \cite{Garcia-Montero:2023gex,Garcia-Montero:2024jev}, or a hadron transport \cite{Schafer:2021csj}. 
It is also possible to study BSQ (baron, strangeness, electric charge) fluctuations where net-$n_B=0$ through gluon splittings into $q\bar{q}$ pairs \cite{Carzon:2019qja}.
These initial states may be far-from-equilibrium and require a pre-equilibrium state incorporating BSQ charges \cite{Carzon:2023zfp,DoreSQM}. 
They are fed into relativistic viscous BSQ hydrodynamic codes, existing open-source codes  are \texttt{CCAKE} \cite{Plumberg:2024leb}, \texttt{MUSIC} \cite{Monnai:2024pvy}, and \texttt{vHLLE} \cite{Karpenko:2013wva}, each have different equations of motion, numerical algorithms, and EOS solvers. 
Stability constraints \cite{Almaalol:2022pjc} provide further constraints on BSQ hydro.
Phase transitions \cite{Kapusta:2024riv} or critical fluctuations \cite{Karthein:2024ipl,An:2022jgc} incorporation in relativistic  hydrodynamics is underway but not yet implemented in dynamical frameworks. 
Hadronization also requires criticality  \cite{Pradeep:2022eil}.
For low $\sqrt{s}$ the influence of the QGP  may  short lived such that it is negligible \cite{Inghirami:2022afu}.  An alternative  is to incorporate potentials within a hadron transport to simulation phase transitions \cite{Sorensen:2020ygf}.


{\bf NS} Scientists can now extract posteriors of the NS EOS ($T=0$, large $n_B$, at $\beta$-equilibrium). 
Different approaches are used: functional forms, physics based EOS (often multiple stitched together in different ranges of $n_B$), or a mixture. 
There is strong evidence that the speed of sound squared surpasses the conformal limit i.e. $c_s^2>1/3$ to support heavy NS \cite{Bedaque:2014sqa} and may near the causal limit if ultra heavy NS exist \cite{Tan:2020ics,Kojo:2020krb}.
A bump in $c_s^2$ could indicate a cross-over  into quarks \cite{McLerran:2018hbz} or a phase transition into hyperons \cite{Cruz-Camacho:2024odu}.
Functional forms should  capture features in $c_s^2$ that indicate phase transitions/new degrees of freedom, which is not possible with polytropes/spectral EOS \cite{Tan:2020ics,Tan:2021ahl}. 
Non-parametric approaches can reproduce these features like Gaussian processes \cite{Essick:2023fso}, these features can also be built in  \cite{Mroczek:2023zxo}.
Strong phase transitions  could be measured via mass twins \cite{Alford:2013aca} or the binary love relation \cite{Tan:2021nat}. 
We cannot calculate the EOS directly at large $n_B$. 
However, lattice QCD calculations are possible at $T=0$ with isospin asymmetry, leading to $c_s^2>1/3$ \cite{Brandt:2022hwy}. 
pQCD calculations are relevant at $n_B\gtrsim 40\,n_{sat}$. 
Using causality and stability, one can obtain bounds that reach to smaller $n_B$ to further constrain the EOS \cite{Komoltsev:2021jzg}.
Bulk viscosity within NS mergers arises from weak interactions, affecting different stages of the merger \cite{Most:2021zvc}, and is influenced by the degrees of freedom \cite{Alford:2020pld} and $Y_Q$ \cite{Yang:2023ogo}. 
In \cite{Ripley:2023lsq} the first constraints on bulk viscous in the inspiral of NS mergers  was extracted, making tests possible with further data.

Groups use low-energy HIC flow data  to extract the low $T\geq 0$ EOS for $Y_Q\sim0.5$.
It is important to consider structure in $c_s^2$, as was done first in \cite{Oliinychenko:2022uvy}, indicating a large bump in $c_s^2$ around $[2,3]n_{sat}$. 
It is possible to convert a given NS EOS into HIC using the symmetry energy expansion (and applying saturation and causality/stability constraints) \cite{Yao:2023yda}, which indicated  ultraheavy NS are consistent with HIC.
Next is to expand cold EOS into finite $T$, which is possible using a $T/\mu_B$ expansion  \cite{Mroczek:2024sfp}  or an effective mass (npe matter) \cite{Raithel:2019gws}.

\section{Bulk Dynamics}
\label{sec:Hadrons}
The QGP is well described by a relativistic viscous fluid with small viscosity \cite{Heinz:2013th}, leading to significant developments of relativistic \emph{viscous} fluids. At $n_B=0$ and large systems, the properties can be exacted from a Bayesian analysis \cite{Bernhard:2019bmu,JETSCAPE:2020mzn,Nijs:2020roc}. 
However,  the QGP medium may begin extremely far-from-equilibrium such that even causality and stability break down \cite{Plumberg:2021bme}, which affects the extraction of bulk viscosity \cite{Domingues:2024pom}. 
Collisions of light nuclei pushes the boundaries of fluid dynamics even further \cite{PHENIX:2018lia}, spin hydrodynamics is relevant for $\Lambda$ polarization results \cite{STAR:2017ckg,Florkowski:2017ruc}, and magnetohydrodynamics  is needed to explore  large magnetic fields \cite{Gursoy:2014aka}.
\emph{Key Questions: } Does HIC knowledge of rel. viscous
fluids affect other fields? Could hydrodynamics be
relevant for the EIC? What role does spin and magnetic fields play?

{\bf Advances in fluid dynamics} Predictions between $\sqrt{s}$ \cite{Niemi:2015voa,Noronha-Hostler:2015uye} were confirmed \cite{ALICE:2016ccg}, so one could  extract nuclear structure information from collisions of deformed ions \cite{Bally:2022vgo}. 
One can extract  the neutron skin from $^{208}Pb$ \cite{Giacalone:2023cet} from HIC, which was more consistent with ab initio \cite{Hu:2021trw} than the PREXII data \cite{PREX:2021umo}.
Collaborations with nuclear structure physicists have led to suggestions of new  run to shed light on $\alpha$ clustering \cite{Summerfield:2021oex} or interesting deformations \cite{Giacalone:2024luz,Giacalone:2024ixe}.
Fundamental questions still remain for small systems and peripheral collisions. Correlations between $v_2$ and $\langle p_T\rangle$ are hard to reproduce \cite{ALICE491854}.
In extremely small systems i.e. a vector meson-ion collisions (ultra peripheral collisions) there are hints of collective behavior \cite{Zhao:2022ayk}, which has interesting implications for the Electron Ion Collider. 
These advanced have led to cold atoms connections \cite{Floerchinger:2021ygn}.

{\bf Spin and Magnetic Fields}
HIC have extremely large, short lived magnetic fields \cite{Inghirami:2019mkc}. Groups are working on developing relativistic magnetohydrodyanmics (MHD) codes for HIC. 
Since ideal MHD is standard in astrophysics, at least two astro codes have been converted to HIC \cite{Inghirami:2016iru,Mayer:2024kkv}.   HIC physicists  and astrophysicists incorporated Israel-Stewart  equations of motion into astro MHD codes \cite{Most:2021rhr}.
Non-linear stability and causality techniques from HIC \cite{Bemfica:2020xym}, have been extended to MHD coupled to general relativity for accretion disks around black holes \cite{Cordeiro:2023ljz}.
Similar constraints are also available for spin/chiral rel. hydrodynamics \cite{Abboud:2023hos,Daher:2024bah}.
$\Lambda$ polarization can further constrain hydrodynamic parameters \cite{PalermoSQM}.

\section{Hadron Physics}
\label{sec:Hadrons}
HIC cannot directly probe quarks and gluons, rather detectors measure  charged hadrons. 
Understanding their properties provides insight into  freeze-out, hadronization, and  Brownian motion.
\emph{Key Questions:} How is strangeness produced?  What causes isospin breaking? Can we understand heavy flavor across system size? 
Could charm quarks be thermalized?

{\bf Puzzles from hadrons} The hadron resonance gas (HRG) model helped understand hadronization.  
HRG is used to extract freeze-out using thermal fits \cite{Andronic:2017pug} and  fluctuations \cite{Bellwied:2018tkc,Alba:2020jir} (see also in lattice QCD \cite{Borsanyi:2014ewa}). 
A tension remains between light and strange hadrons (see \cite{Floris:2014pta}).
Possibly increasing the number of resonances would solve the puzzle \cite{Bazavov:2014xya}, but the tension remained even with new states \cite{Alba:2020jir,SanMartin:2023zhv}.
Either there are separate freeze-out temperatures  \cite{Bellwied:2013cta} or an S-matrix approach is required \cite{Andronic:2018qqt}. 
Extensions to the HRG include magnetic fields \cite{Vovchenko:2024wbg} or surface tension \cite{ZherebtsovaSQM}.
Experiments found isospin symmetry breaking of kaons across $\sqrt{s}$ \cite{Brylinski:2023nrb}. 
One explanation  is  a disorientated isospin condensate \cite{Kapusta:2023xrw}.

{\bf Puzzles from heavy flavor} Heavy flavor  provides an interesting probe of the QGP that is normally modeled through a Langevin equation at low  $p_T$, or an energy loss model at high $p_T$.
Most dynamics are described using more weakly coupled approaches \cite{BassSQM}, but others have used charm as a conserved charge within the fluid \cite{Capellino:2022nvf,Capellino:2023cxe}.
Coalescence is important at low $p_T$  to correctly capture the hadronization process (see \cite{HirayamaSQM} for hadronic rescattering). 
While it was thought heavy flavor models lead to disparate results for the nuclear modification factors $R_{AA}$,  much could be attributed to different medium effects since the same medium led to very similar results \cite{Cao:2018ews}. 
After the first event-by-event studies with heavy flavor \cite{Nahrgang:2014vza} and jets \cite{Noronha-Hostler:2016eow}, soft-heavy \cite{Prado:2016szr,Sambataro:2022sns} and soft-hard \cite{Barreto:2022ulg,Holtermann:2023vwr} correlations advanced. 
These correlations can constrain the $T$ dependence of diffusion \cite{Sambataro:2023tlv}.
ALICE experimental data showed that the heavy flavor elliptical flow, $v_2$, scales with the soft $v_2$ \cite{Sambataro:2022sns}. 
Charmonium flow may be more complicated due to its internal structure \cite{Cho:2023kpe}.
Jets' sensitivity to event-by-event fluctuations can be seen by experimental measurements of jet $v_3\neq =0$ \cite{ATLAS:2021ktw}.
Groups are trying to unify the soft-heavy \cite{Zhao:2024ecc} and soft-hard \cite{Karmakar:2024jak} frameworks.

Small systems have  perplexing behavior for  heavy flavor. $R_{AA}\sim 1$, while  pPb D meson $v_2$ is large  
\cite{CMS:2018loe} but charmonium \cite{CMS:2018duw} and bottomonium $v_2\sim 0$ \cite{CMS:2023dse} and in pp D mesons  $v_2>0$  and B mesons $v_2<0$ \cite{ATLAS:2023vms}.
A heavy flavor system size scan was proposed \cite{Katz:2019fkc}  with  multicharm hadrons predictions in \cite{Minissale:2023dct} . 
Corrections to the short path length may be required \cite{Faraday:2023mmx}.
Heavy flavor requires longer formation times \cite{Moore:2004tg} due to their masses. 
The interplay between the early far-from-equilibrium stage and heavy flavor is intriguing, such as how $c\bar{c}$ and $b\bar{b}$ pairs disassociate in the glasma \cite{Pooja:2024rnn}.

{\bf Acknowledgments} J.N.H thanks  SQM24 for providing child care and acknowledges support from DOE Grant No. DE-SC0023861,  Saturated Glue (SURGE) Topical Theory Collaboration, and NSF Grant No. OAC-2103680  within the framework of MUSES collaboration.

\bibliography{inspire,NOTinspire}

\end{document}